\documentstyle[12pt,amssymb,epsfig]{article}
\def\theequation{\arabic{section}.\arabic{equation}}

\begin{document}

\def\tr{\mathop{\rm tr}}
\def\Tr{\mathop{\rm Tr}}
\def\Im{\mathop{\rm Im}}
\def\Re{\mathop{\rm Re}}
\def\bR{\mathop{\bf R}{}}
\def\bC{\mathop{\bf C}{}}
\def\C{\mathop{\rm C}}
\def\bra#1{\left\langle #1\right|}
\def\ket#1{\left| #1\right\rangle}
\def\VEV#1{\left\langle #1\right\rangle}
\def\gdot#1{\rlap{$#1$}/}
\def\abs#1{\left| #1\right|}
  \newcommand{\ccaption}[2]{
    \begin{center}
    \parbox{0.85\textwidth}{
      \caption[#1]{\small{\it{#2}}}
      }
    \end{center}
    }
\def\beq{\begin{equation}}
\def\eeq{\end{equation}}
\def\eq{\beq\eeq}
\def\beqn{\begin{eqnarray}}
\def\eeqn{\end{eqnarray}}
\relax
\let\h=\hat
\newcommand\SS{\scriptsize}
\newcommand\sss{\scriptscriptstyle}
\newcommand\gs{g_{\sss S}}
\newcommand\as{\alpha_{\sss S}}         
\newcommand\ep{\epsilon}
\newcommand\Th{\theta}
\newcommand\epb{\overline{\epsilon}}
\newcommand\aem{\alpha_{\rm em}}
\newcommand\refq[1]{$^{[#1]}$}
\newcommand\avr[1]{\left\langle #1 \right\rangle}
\newcommand\lambdamsb{\Lambda_5^{\rm \sss \overline{MS}}}
\newcommand\qqb{{q\bar{q}}}
\newcommand\qb{\bar{q}}
\newcommand\xto{\tilde{x}_1}
\newcommand\xtt{\tilde{x}_2}
\newcommand\szeroi{{\cal S}_i^{(0)}}
\newcommand\aoat{a_1 a_2}
\newcommand\Jetlist{\{J_l\}_{1,3}}
\newcommand\SFfull{\{k_l\}_{1,6}}
\newcommand\SFfullbi{\{k_l\}_{1,6}^{[i]}}
\newcommand\SFTfull{\{k_l\}_{1,5}}
\newcommand\SFj{\{k_l\}_{3,6}}
\newcommand\FLfull{\{a_l\}_{1,6}}
\newcommand\FLfullbi{\{a_l\}_{1,6}^{[i]}}
\newcommand\FLTfull{\{a_l\}_{1,5}}
\newcommand\FLFj{\{a_l\}_{3,6}}
\newcommand\SFjbi{\{k_l\}_{3,6}^{[i]}}
\newcommand\FLjbi{\{a_l\}_{3,6}^{[i]}}
\newcommand\SFjexcl{\{k_l\}_{i,j}^{[np..]}}
\newcommand\SCollfull{\{k_l\}_{1,7}^{[ij]}}
\newcommand\SColl{\{k_l\}_{3,7}^{[ij]}}
\newcommand\FLCollfull{\{a_l\}_{1,7}^{[ij]}}
\newcommand\FLColl{\{a_l\}_{3,7}^{[ij]}}
\newcommand\STj{\{k_l\}_{3,5}}
\newcommand\FLTj{\{a_l\}_{3,5}}
\newcommand\FLNj{\{a_l\}_{3}^{N+2}}
\newcommand\FLNmoj{\{a_l\}_{3}^{N+1}}
\newcommand\FLNfullj{\{a_l\}_{1}^{N+2}}
\newcommand\FLNmofullj{\{a_l\}_{1}^{N+1}}
\newcommand\Argfull{\FLfull;\SFfull}
\newcommand\ArgTfull{\FLTfull;\SFTfull}
\newcommand\KtoKF{(k_1,k_2\to\SFj\,;\FLFj)}
\newcommand\KtoKT{(k_1,k_2\to\STj\,;\FLTj)}
\newcommand\FLsum{\sum_{\{a_l\}}}
\newcommand\FLFsum{\sum_{\FLFj}}
\newcommand\FLFsumbi{\sum_{\FLjbi}}
\newcommand\FLTsum{\sum_{\FLTj}}
\newcommand\FLNsum{\sum_{\FLNj}}
\newcommand\FLNmosum{\sum_{\FLNmoj}}
\newcommand\MF{{\cal M}^{(4)}}
\newcommand\MT{{\cal M}^{(3)}}
\newcommand\MTz{{\cal M}^{(3,0)}}
\newcommand\MTo{{\cal M}^{(3,1)}}
\newcommand\MTi{{\cal M}^{(3,i)}}
\newcommand\MTmn{{\cal M}^{(3,0)}_{mn}}
\newcommand\MN{{\cal M}^{(N)}}
\newcommand\MNmo{{\cal M}^{(N-1)}}
\newcommand\MNmoij{{\cal M}^{(N-1)}_{ij}}
\newcommand\MNmoV{{\cal M}^{(N-1,v)}}
\newcommand\MFsj{\MF(k_1,k_2\to\SFj)}
\newcommand\MTsj{\MT(k_1,k_2\to\STj)}
\newcommand\MTisj{\MTi(k_1,k_2\to\STj)}
\newcommand\PHIFsj{\phi_4(k_1,k_2\to\SFj)}
\newcommand\PHITsj{\phi_3(k_1,k_2\to\STj)}
\newcommand\uoffct{\frac{1}{4!}}
\newcommand\uotfct{\frac{1}{3!}}
\newcommand\uoNfct{\frac{1}{N!}}
\newcommand\uoNmofct{\frac{1}{(N-1)!}}
\newcommand\uoxic{\left(\frac{1}{\xi}\right)_c}
\newcommand\uoxiic{\left(\frac{1}{\xi_i}\right)_c}
\newcommand\uoxilc{\left(\frac{\log\xi}{\xi}\right)_c}
\newcommand\uoxiilc{\left(\frac{\log\xi_i}{\xi_i}\right)_c}
\newcommand\uoyim{\left(\frac{1}{1-y_i}\right)_+}
\newcommand\uoyimdi{\left(\frac{1}{1-y_i}\right)_{\delta_{\sss I}}}
\newcommand\uoyimpdi{\left(\frac{1}{1\mp y_i}\right)_{\delta_{\sss I}}}
\newcommand\uoyjmdo{\left(\frac{1}{1-y_j}\right)_{\delta_o}}
\newcommand\uoyip{\left(\frac{1}{1+y_i}\right)_+}
\newcommand\uoyipdi{\left(\frac{1}{1+y_i}\right)_{\delta_{\sss I}}}
\newcommand\uoyilm{\left(\frac{\log(1-y_i)}{1-y_i}\right)_+}
\newcommand\uoyilp{\left(\frac{\log(1+y_i)}{1+y_i}\right)_+}
\newcommand\uozm{\left(\frac{1}{1-z}\right)_+}
\newcommand\uozlm{\left(\frac{\log(1-z)}{1-z}\right)_+}
\newcommand\SVfact{\frac{(4\pi)^\ep}{\Gamma(1-\ep)}
                   \left(\frac{\mu^2}{Q^2}\right)^\ep}
\newcommand\Oop{{\cal O}}
\newcommand\Sfun{{\cal S}}
\newcommand\Pfun{{\cal P}}
\newcommand\MSB{{\rm \overline{MS}}}
\newcommand\DIG{{\rm DIS}_\gamma}
\newcommand\CA{C_{\sss A}}
\newcommand\DA{D_{\sss A}}
\newcommand\CF{C_{\sss F}}
\newcommand\TF{T_{\sss F}}


\begin{titlepage}
\nopagebreak
{\flushright{
        \begin{minipage}{4cm}
        CERN-TH/98-255 \hfill \\
        ETH-TH/98-21 \hfill \\
        hep-ph/9808262\hfill \\
        \end{minipage}        }

}
\vfill
\begin{center}

{\large {\sc Next-to-Leading Order Jet Cross Sections\\
             in Polarized Hadronic Collisions \\ }} 
\vskip .5cm
{\bf Daniel de Florian}
\\                    
\vskip .1cm
{CERN, TH Division, Geneva, Switzerland}\\
\vskip .5cm
{\bf Stefano Frixione}\footnote{Work supported by the Swiss National
Foundation.}
\\                    
\vskip .1cm
{Theoretical Physics, ETH, Zurich, Switzerland} \\
\vskip .5cm
{\bf Adrian Signer and Werner Vogelsang}
\\                    
\vskip .1cm
{CERN, TH Division, Geneva, Switzerland}\\
\end{center}
\nopagebreak
\vfill
\begin{abstract}
We present a next-to-leading order computation in QCD of one-jet and
two-jet cross sections in polarized hadronic collisions. Our results
are obtained in the framework of a general formalism that deals with
soft and collinear singularities using the subtraction method.
We construct a Monte Carlo program that generates events at the
partonic level. We use this code to give phenomenological predictions
for $pp$ collisions at $\sqrt{S}=500$~GeV, relevant for the spin 
physics program at RHIC. The possibility of using jet data to constrain
the poorly known polarized parton densities is examined.
\end{abstract}        
\vfill
\end{titlepage}

\section{Introduction}
\setcounter{equation}{0}

In the last few years, measurements \cite{data} of the spin asymmetries 
$A_1^N$ in deep-inelastic scattering (DIS) of longitudinally
polarized lepton beams off polarized hadron ($N=p,n,d$) targets have
provided much new information on the spin structure of the
nucleon. Theoretical leading order (LO) [2--6]
and next-to-leading order (NLO) [2--7] 
analyses of the data sets demonstrate, however,  
that these are not sufficient to accurately extract the spin-dependent quark 
($\Delta q = q^{\uparrow}-q^{\downarrow}$) and gluon 
($\Delta g=g^{\uparrow}-g^{\downarrow}$) densities of the nucleon. 
This is true in particular for $\Delta g(x,Q^2)$, since this quantity
contributes to DIS in LO only via the $Q^2$-dependence of $A_1^N$, which
could not be accurately studied experimentally so far. As a result of
this, it turns out [2--6]
that the $x$-shape of $\Delta g$ seems to be hardly  
constrained at all by the DIS data, even though a tendency towards a 
fairly large positive {\em total} gluon polarization, $\int_0^1 \Delta 
g(x,Q^2=4 \; \mbox{GeV}^2) dx \gtrsim 1$, was found \cite{grsv,gs,dss,bfr}. 
The measurement of $\Delta g$ thus remains one of the most interesting 
challenges for future high-energy experiments with polarized nucleons. 
In selecting suitable processes for a determination of $\Delta g$,
it is crucial to pick those that, unlike DIS, have a direct gluonic
contribution already at the lowest order. Here, one thinks in the first 
place of high-$p_{\sss T}$ reactions in nucleon--nucleon collisions, which
have been tremendously important in the unpolarized case to constrain
the unpolarized gluon density. 

At the moment, the most eagerly awaited experimental tool for the `spin 
physics' community is the RHIC collider at BNL, at which first runs in a 
proton--proton mode with polarized beams are expected to be
accomplished in about two years from now. The centre-of-mass energy
for these $pp$ collisions will be ranging between 100 and 500 GeV,
with luminosities (rising with energy) between 240 and 800 pb$^{-1}$,
respectively. One expects about $70 \%$ polarization for each beam. 
Such conditions look extremely favourable for studying the spin asymmetries
for all kinds of high-$p_{\sss T}$ $pp$ processes that are sensitive to 
the gluon density, such as jet, prompt-photon, or heavy-flavour production.
Considering the higher end of RHIC energies, jets could be {\em the} key
to $\Delta g$: at $\sqrt{S} =500$~GeV, clearly structured jets will be
extremely copiously produced, and jet-observables will show a strong 
sensitivity to $\Delta g$ thanks to the dominance~\cite{soffer} of the 
$gg$ and $qg$ initiated subprocesses in some kinematical ranges.

In order to make reliable quantitative predictions for a high-energy process,
it is crucial to determine the NLO QCD corrections
to the Born approximation. Quite in general, the key issue here is to check
the perturbative stability of the process considered, i.e. to examine
the extent to which the NLO corrections affect the cross sections and
(in spin physics) the spin asymmetries relevant for experimental measurements.
Only if the corrections are under control can a process that shows 
good sensitivity to, say, $\Delta g$ at the lowest order be regarded 
as a genuine probe of the polarized gluon distribution and be reliably 
used to extract it from future data.

NLO QCD corrections are expected to be particularly important for the
case of jet-production, since it is only at NLO that the QCD structure 
of the jet starts to play a r\^{o}le in the theoretical description, providing 
for the first time the possibility to realistically match the experimental 
conditions imposed to define a jet.

The calculation of the NLO QCD corrections to jet production by polarized
hadrons is the purpose of this paper. Such a calculation needs the
one-loop $2\to 2$ and tree-level $2\to 3$ polarized (i.e. {\it not} summed
over external helicities) amplitudes as input. Fortunately, these
amplitudes are already known \cite{MatEl,wu}. Furthermore, several
independent methods to calculate any infrared-safe quantity in any kind of
hard unpolarized collision are at present available in the  
literature~\cite{GGK,FKS,CS}. The formalism of ref.~\cite{FKS}
has been used in ref.~\cite{Jets97} to construct a Monte Carlo code
that can calculate any three-parton infrared-safe observable
in hadron--hadron unpolarized collisions. In the present paper, we will
extend the method of refs.~\cite{FKS,Jets97} and adapt the Monte Carlo 
code to the case of polarized hadron--hadron collisions. As a result, we will 
present a customized code, with which it will be possible to calculate 
any infrared-safe quantity corresponding to either single- or di-jet 
production to NLO accuracy.

The outline of the paper is as follows: In section~\ref{formalism} we
describe the calculation of next-to-leading order corrections to jet cross
sections in polarized hadronic collisions. Since there is no conceptual
difference to the unpolarized case, we restrain from repeating all details
and only give an overview. A crucial input for the next-to-leading order
calculation is the polarized gluon distribution function. Section~\ref{pdf}
summarizes the current situation and discusses the main assumptions that
are used to constrain the various fits. We begin our phenomenological
study in section~\ref{PertStab}, with the investigation of the perturbative
stability and reliability of our next-to-leading order results for 
single-inclusive as well as double-differential
observables. In section~\ref{DepPDF} we turn to the more phenomenological 
issue of the dependence of the cross sections on the various
parametrizations of the parton densities. In particular, we will also
investigate some spin asymmetries. Finally, in
section~\ref{conclusions} we present our conclusions. Some technical
details concerning the difference 
between the calculations of the cross section in polarized and unpolarized
hadronic collisions are given in an appendix.

\section{Formalism \label{formalism}}
\setcounter{equation}{0}

We start by writing a generic differential jet cross section in polarized
hadronic collisions using the factorization theorem~\cite{CSS}
\beqn
d\sigma^{(H_1 H_2)}(K_1,\Lambda_1;K_2,\Lambda_2;\Sfun)&=&
\sum_{\aoat}\sum_{\lambda_1\lambda_2}\int dx_1\,dx_2\,
f^{(H_1\Lambda_1)}_{a_1\lambda_1}(x_1)
f^{(H_2\Lambda_2)}_{a_2\lambda_2}(x_2)
\nonumber \\*&&\phantom{\sum_{\aoat}\sum_{\lambda_1\lambda_2}}\times
d\hat{\sigma}_{\aoat}(x_1 K_1,\lambda_1;x_2 K_2,\lambda_2;\Sfun),
\label{factth1}
\eeqn
where $H_1$ and $H_2$ are the incoming hadrons, with momenta $K_1$ and
$K_2$ and helicities $\Lambda_1$ and $\Lambda_2$ respectively;
$f^{(H_i\Lambda_i)}_{a_i\lambda_i}$ is the  
non-calculable but universal distribution function for the parton $a_i$ 
with helicity $\lambda_i$ in the hadron $H_i$ with helicity $\Lambda_i$ 
and $d\hat{\sigma}_{\aoat}$ are the (subtracted) short-distance partonic 
cross sections\footnote{In the following, we will often omit some of the
  entries of the quantity $d\hat{\sigma}_{\aoat}$; the meaning should be
  clear from the context.}. The quantity $\Sfun$ is the measurement
function; it defines the jet momenta in terms of the parton momenta. Its
specific form therefore depends upon the jet-clustering algorithm adopted.
Since the results for polarized scattering are usually 
presented in terms of the quantity
\beqn 
d\Delta\sigma^{(H_1 H_2)}&=&
\frac{1}{4}\Big(\,\, d\sigma^{(H_1 H_2)}(+;+)+d\sigma^{(H_1 H_2)}(-;-) 
\nonumber \\*&&\,\,\,\, 
-\,d\sigma^{(H_1 H_2)}(+;-)-d\sigma^{(H_1 H_2)}(-;+)\Big),
\label{Deltadef}
\eeqn
it is convenient to rewrite eq.~(\ref{factth1}) in the following form:
\beq
d\Delta\sigma^{(H_1 H_2)}(K_1;K_2;\Sfun)=\sum_{\aoat}\int dx_1\,dx_2\,
\Delta f^{(H_1)}_{a_1}(x_1)\Delta f^{(H_2)}_{a_2}(x_2)
d\Delta\hat{\sigma}_{\aoat}(x_1 K_1;x_2 K_2;\Sfun)\,,
\label{factth2}
\eeq
where $d\Delta\hat{\sigma}_{\aoat}$ is defined analogously to
$d\Delta\sigma^{(H_1 H_2)}$ in eq.~(\ref{Deltadef}), and
\beq
\Delta f^{(H_i)}_{a_i}=f^{(H_i+)}_{a_i+}-f^{(H_i+)}_{a_i-}
=f^{(H_i-)}_{a_i-}-f^{(H_i-)}_{a_i+}.
\eeq
The available parametrizations for polarized parton densities are
presented in terms of the quantities $\Delta f^{(H)}_{a}$.

Our aim in the present paper is to evaluate the polarized cross
section in eq.~(\ref{factth2}) at next-to-leading order in
perturbative QCD, for one- and two-jet production. As is well
known, there are two main problems in such a computation. First of all,
the relevant polarized amplitudes have to be
known. For our case, we need the $2\to 2$ one-loop and $2\to 3$ tree-level
amplitudes. These amplitudes have been computed \cite{MatEl,wu} using the
technique of colour ordering and the helicity method. Secondly, the
cancellation of the infrared poles, which appear in the intermediate steps
of the calculation, has to be performed analytically. Due to the universal
(i.e. process-independent) structure of these poles, general
methods~\cite{GGK,FKS,CS} exist, which allow the computation 
of any infrared-safe cross section in any type of hard scattering.
In the present paper we will use the approach of ref.~\cite{FKS},
based upon the subtraction method. The main idea of ref.~\cite{FKS}
is to exploit the properties of the measurement function in order
to disentangle the infrared-singular regions that appear in the
real contribution. Indeed, as follows from the universal properties
of the measurement function (which are responsible for the good definition
of any infrared-safe cross section), this quantity can be written
as a sum of terms, each of which is non-vanishing only when one given
parton is soft or collinear to another parton. Therefore, the
partonic cross section can be expressed as a sum of terms whose
singular structure is trivial, and the subtraction procedure can
be straightforwardly implemented. It is important to notice that, although
the approach of ref.~\cite{FKS} was originally introduced in the
case of unpolarized collisions, it does not need any principle 
modification to be applied to the case of polarized collisions,
since the major r\^{o}le in the treatment of the singularities is
played by the measurement function. Further details on this topic
can be found in the appendix.

The procedure of ref.~\cite{FKS} results in the subtracted partonic
cross sections which appear in eqs.~(\ref{factth1}) or~(\ref{factth2}).
Standard Monte Carlo methods can therefore be used to compute 
the quantity
\beq
\langle H \rangle =\int\,d\Delta\sigma^{(H_1 H_2)}(\Sfun)\,H,
\label{Hint}
\eeq
where $H$ is any function of the jet momenta, which are defined by 
$\Sfun$. If H is a product of $\theta$ functions, implementing experimental
cuts and selecting a bin of a given histogram, then $\langle H \rangle$ is
the QCD prediction for the cross section in that bin.

The main drawback of eq.~(\ref{Hint}) is that the jet definition
is used, through the measurement function, to disentangle the 
infrared singularities. This prevents us from getting predictions
for different jet definitions when performing a single computer
run, as is customary in a parton shower Monte Carlo; also, one-jet
and two-jet observables must be treated separately. This problem
was addressed in ref.~\cite{Jets97}; the key observation is that,
in order to disentangle the infrared singularities, the measurement
function can be substituted by a suitable sum of products of $\theta$ 
functions, which we call the $\Pfun$ function. By construction, 
each term of the sum is non-vanishing only in one
given infrared-singular region which contributes to the cross
section at next-to-leading order. In this sense, the $\Pfun$ function
and the $\Sfun$ function are completely equivalent, and the former
can be interpreted as a fake measurement function (however, while
the real measurement function contains $\delta$ functions that define
the jet momenta in terms of the parton momenta, the $\Pfun$ function
only contains $\theta$ functions). Therefore, following ref.~\cite{Jets97}, 
we can write the analogue of eq.~(\ref{Hint}) as 
\beq
[H\Sfun]=\int\,d\Delta\sigma^{(H_1 H_2)}(\Pfun)\,H\,\Sfun.
\label{HSint}
\eeq
By construction, we get \mbox{$\langle H \rangle =[H\Sfun]$}. A computer
program based upon eq.~(\ref{Hint}) outputs jet momenta, which are
eventually used to fill a histogram as specified by $H$. On the other hand,
a computer program based upon eq.~(\ref{HSint}) outputs parton momenta;
these quantities are eventually used to compute jet momenta (as specified 
by $\Sfun$), which will again fill the histogram given by $H$. Therefore, 
in eq.~(\ref{HSint}) no jet definition is involved in the generation of the
hard event and in the computation of the weight. It follows that,
during the same computer run, it is possible to use several different
jet definitions and to compute one-jet, two-jet and non-jet (like
transverse thrust) observables. This kind of computer code is called a
{\it parton generator}.

In ref.~\cite{Jets97} two parton generators were presented, one for
photon--hadron collisions and one for hadron--hadron collisions. The
latter has been suitably modified to deal with polarized hadron--hadron
collisions, and used to produce the phenomenological results presented
in this paper. The structure of the code of ref.~\cite{Jets97} remains
unchanged, since only the unpolarized partonic cross sections and splitting 
functions had to be substituted with the polarized ones (see the appendix 
for more details).

\section{Polarized Parton Distribution Functions \label{pdf}}
\setcounter{equation}{0}

As stated in the introduction, there is hardly any experimental information 
on the spin-dependent gluon density $\Delta g$ at present. In contrast, 
the quark densities are far better constrained by the existing data from 
inclusive polarized DIS. This is particularly true for the polarized valence 
quark densities, which come out rather similar in all theoretical analyses 
performed so far. The spin-dependent sea-quark distributions seem less well
constrained; however, they are of minor importance for our jet 
studies\footnote{Strictly speaking, inclusive DIS data for proton and 
neutron targets can only give information on the two non-singlet and
the quark singlet combinations, rather than on all quark and antiquark 
densities individually. The inclusion of data~\cite{data} for the 
spin asymmetry in semi-inclusive DIS (SIDIS), which in principle would allow a 
complete flavour discrimination in the quark sector, does not provide much 
help in practice, since the SIDIS data have not yet reached the 
precision of the inclusive ones. The flavour decomposition of the nucleon
parton densities -- which is essential for making predictions for observables 
other than DIS -- therefore partly depends on theoretical assumptions
made when analysing the DIS data~[2--5].}.

\begin{figure}
\centerline{
   \epsfig{figure=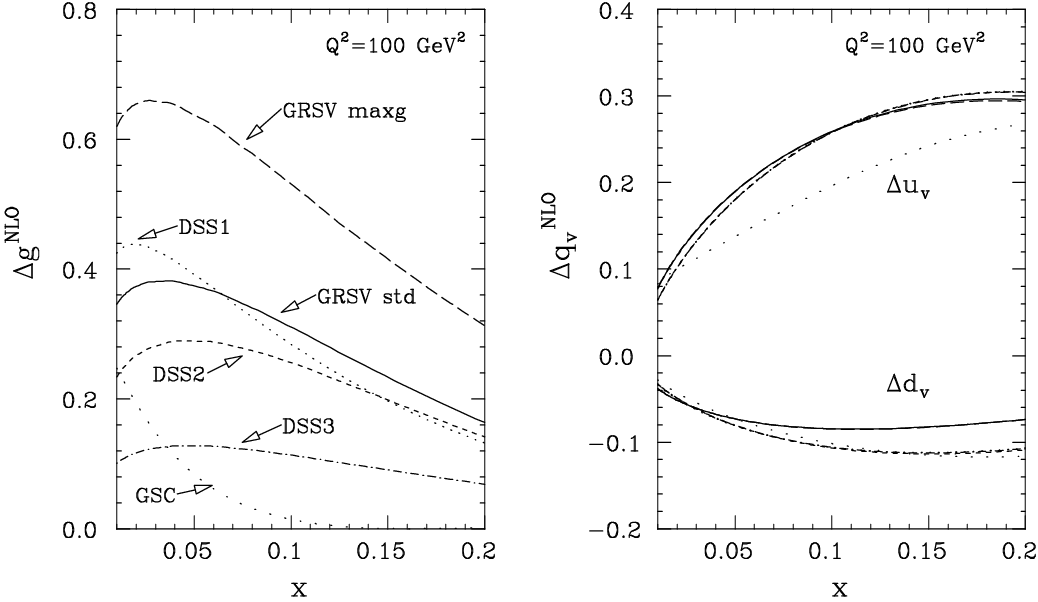,width=0.64\textwidth,clip=} }
\ccaption{}{ \label{figPDF}
   The polarized gluon (left) and valence quark densities (right), as
   given by the six NLO parametrizations that will be used in this paper,
   at the scale $Q^2 = 100\ \mathrm{GeV}^2$. The patterns for the quark
   densities are the same as those used for the gluon.
}
\end{figure}                                                              

In our phenomenological analysis we will try to cover as much as possible 
of the wide range of polarized parton densities allowed by the present DIS  
data: this is especially relevant for the gluon density, as we will
discuss below. We will use the following six polarized parton density 
sets, which were obtained within various theoretical analyses of 
polarized DIS:
\begin{itemize}
\item
Two sets of ref.~\cite{grsv}, the `standard' set, corresponding to the best 
fit to the data obtained in ref.~\cite{grsv} (from now on, 
referred to as GRSV std),
and a set obtained when saturating the positivity constraint 
$|\Delta g|\leq g$ for the gluon density at the low input scale 
${\cal{O}}(0.5$ GeV) (GRSV maxg). 
\item
The three sets of ref.~\cite{dss} (DSS1, DSS2, DSS3), obtained by 
constraining the first moment of the polarized gluon densities in three
different ways. The first moment of the gluon of DSS1 is the largest and
almost one order of magnitude larger than for DSS3, which has the smallest 
gluon. 
\item
Set C of ref.~\cite{gs} (GS-C), which provides a gluon distribution 
with a qualitatively different $x$-shape, becoming negative at large $x$
for low $Q^2$. Sets A and B of ref.~\cite{gs} are similar to some of the
GRSV and DSS ones, and we do not use them so as to avoid a proliferation
of curves. 
\end{itemize}
All these distributions are available at both LO and NLO, the latter 
corresponding to the $\overline{{\mbox{MS}}}$ scheme used also in our 
calculation of the NLO partonic cross section.

In this work, we will analyse the phenomenology of polarized jet production 
for the particular case of RHIC at a centre-of-mass energy of $\sqrt{S}=500$ 
GeV and for jet transverse momenta in the region of 
$15 \ {\rm GeV} < p_{\sss T}< 100$ GeV. This implies that the polarized
parton distributions will mainly be probed in  the $x$-range
$0.06 \lesssim x \lesssim 0.4$ and at typical scales  $Q^2$ of the order
of 100 GeV$^2$. Figure~\ref{figPDF} shows the NLO polarized valence quark
and gluon densities of the six different sets we are going to use. It
becomes obvious that there is indeed a wide range of possible gluon
distributions compatible with present polarized DIS data. As mentioned
earlier, in the valence sector, most of the distributions are very
similar, with slight exceptions in the cases of the $u_v$ distribution of
GS-C and the $d_v$ density of GRSV std. The origin of these differences
can be easily traced back to the fact that the analyses of
refs.~\cite{grsv,gs} are based on a somewhat smaller data sample than the
others, since some data sets became available only after~\cite{grsv,gs}
had been published. The spin-dependent sea-quark densities (not shown in
fig.~\ref{figPDF}) differ more strongly among the various sets, but have
only a very small impact on the jet cross sections in this kinematical
region: they only amount to less 
than $5\%$ of the contribution from valence quarks. In conclusion, since the
variations in the quark sector are much smaller than the ones for gluons, 
we can expect that any differences between predictions for the polarized jet 
cross sections (or asymmetries) that are found when using different 
polarized parton density sets, are to be attributed to the 
sensitivity of the observable to $\Delta g$.

\begin{figure}
\centerline{
   \epsfig{figure=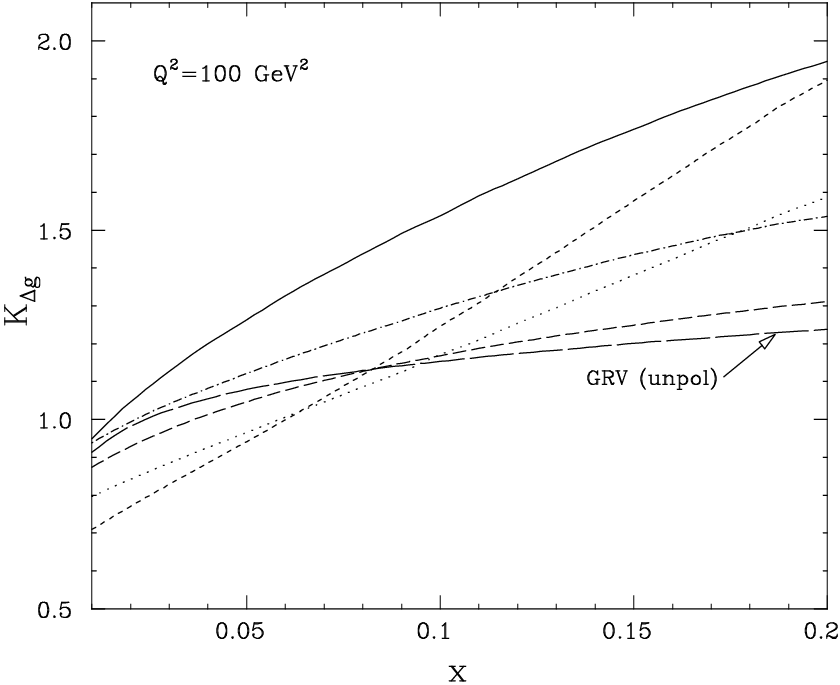,width=0.48\textwidth,clip=} }
\ccaption{}{ \label{figKGLUE}
   Ratio of NLO to LO polarized gluon densities. We use the same pattern as
   in fig.~\ref{figPDF} to distinguish the various lines. Also shown is
   the ratio for an unpolarized (GRV) set.  
}
\end{figure}                                                              

The size of radiative QCD corrections to a given unpolarized hadronic
process is often displayed in terms of a `$K$-factor' which represents the
ratio of the NLO over LO results. In the calculation of the numerator of
$K$ one obviously has to use NLO-evolved parton densities. As far as the
denominator is concerned, a natural definition requires the use of
LO-evolved parton densities. However, by using NLO-evolved parton
densities and LO partonic cross sections, one still obtains
a hadronic cross section accurate to LO, and therefore the
denominator of the $K$-factor can also be computed with
NLO-evolved parton densities. If one chooses a `natural'
subtraction scheme, such as $\overline{{\rm MS}}$,
these two definitions of the $K$-factor are expected to give
similar results (we stress that the two definitions might give
rather different results in the framework of an arbitrary subtraction
scheme: there is no reason to worry about that, since the
$K$-factor is {\it not} a physical quantity, and therefore it
is not supposed to be scheme-independent). However, in the
case of polarized scattering, additional problems arise.
Indeed,  suppose one attempts to fit $\Delta g$ from the 
DIS data. Since the data hardly constrain the gluon density, very different 
results for $\Delta g$ can emerge if the fit is performed at LO or at NLO.
This is confirmed in fig.~\ref{figKGLUE}, where we show the `gluonic' 
$K$-factors  $K_{\Delta g} \equiv \Delta g^{\rm NLO}/\Delta g^{\rm LO}$ as
functions of $x$ for our various sets\footnote{We do not show $K_{\Delta
    g}$ for GS-C, since in this case $\Delta g^{\rm LO}$ may be zero.}.
It can be seen that indeed for most sets $K_{\Delta g}$ is {\em not} close to
unity. We observe that things are much better in the unpolarized case,
where there are far more data to constrain the gluon: here $K_g \equiv
g^{\rm NLO}/g^{\rm LO} \approx 1$. Also note that in the case of the `GRSV
maxg' set,  where one assumes~\cite{grsv} 
that $\Delta g^{\rm LO,NLO}=g^{\rm LO,NLO}$ at the input scale 
(see above) in both LO and NLO, a $K_{\Delta g} \approx 1$ is found also 
at higher scales. This underlines our point that artificially large or small 
$K$-factors for, say, polarized jet production could result merely 
from the fact that the gluon is at present so ill-constrained. We will
discuss this point further in section~\ref{PertStab}.

\section{Perturbative Stability \label{PertStab}}
\setcounter{equation}{0}

\begin{figure}
\centerline{
   \epsfig{figure=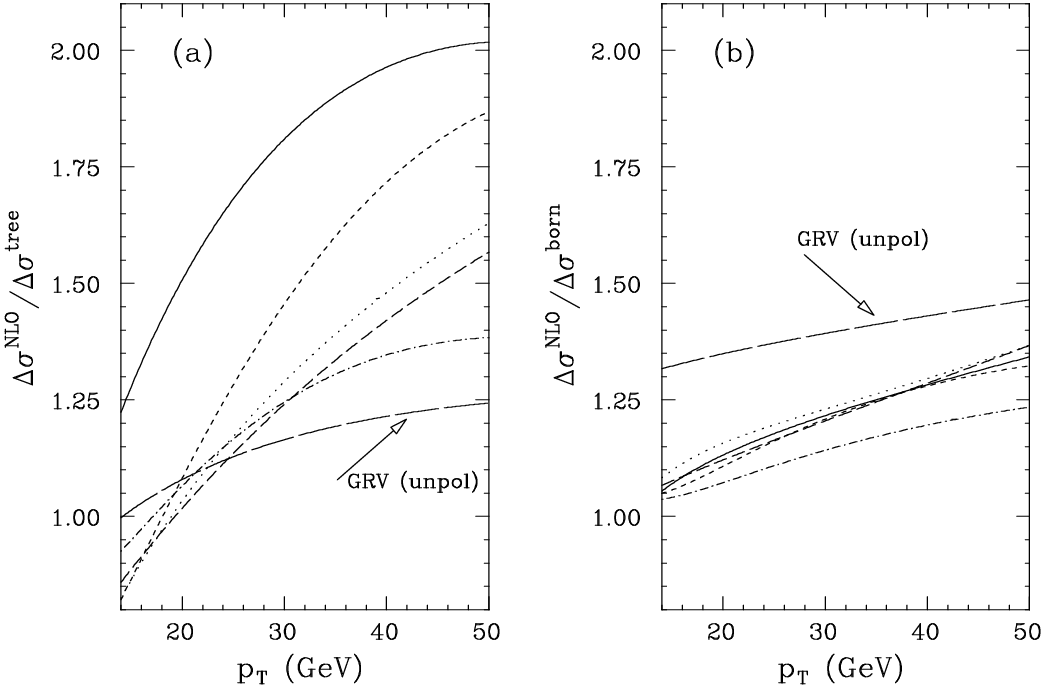,width=0.64\textwidth,clip=} }
\ccaption{}{ \label{figLO_BORN}
Ratio of the next-to-leading order cross section over (a) the tree-level
   cross section (i.e. leading order pdf) and (b) the Born cross section
   (i.e. next-to-leading order pdf) for various parton densities. We use
   the same pattern as in fig.~\ref{figPDF} to distinguish the various
   parton densities.  
}
\end{figure}

As we mentioned in the previous section, in the case of polarized
collisions the study of the $K$-factor, which in general does not
give any information on the perturbative stability of the results
in hadronic scattering, faces additional problems, due to the
large difference between $\Delta g^{\rm NLO}$ and $\Delta g^{\rm LO}$.
To illustrate this issue, we have computed the LO hadronic cross section
entering the $K$-factor in two different ways, by convoluting the
partonic LO cross sections with either LO-evolved parton densities or
NLO-evolved parton densities. For the sole purpose of distinguishing 
the two definitions, we will call the former the tree-level cross section 
and the latter the Born cross section. As an example we
have chosen the $p_{\sss T}$ spectrum, using the Ellis--Soper (ES) cluster
jet algorithm as proposed in ref.~\cite{EScluster} with the resolution
parameter $D=1$. In fig.~\ref{figLO_BORN}a we show the ratio of the NLO
cross section over the tree-level cross section. This ratio can be rather
large for some parton density sets. However, a comparison with
fig.~\ref{figKGLUE} shows that these large corrections come mainly from
the change of $\Delta g^{\rm LO}$ to $\Delta g^{\rm NLO}$. Indeed, if we
plot the ratio of the NLO cross section over the Born cross section, as
done in fig.~\ref{figLO_BORN}b, we see that the corrections are
moderate. For the same reason as in fig.~\ref{figKGLUE} we do not show the
curve for the GS-C set. On the other hand, we also show the same ratio for
the unpolarized case, using the GRV parton densities \cite{grv}.

A reliable error estimate on our NLO results requires some knowledge on the
size of the uncalculated higher-order terms. The only fully reliable way
to get this information is to perform computations of even higher order. 
Unfortunately, such a calculation is currently out of reach. Thus, the
best we can do is to study  the dependence of the full NLO  results
on the renormalization and factorization scales. Throughout we will set
the two scales equal, i.e. $\mu_R = \mu_F \equiv \mu$. Although physical
observables are obviously independent of $\mu$, theoretical predictions do
have such a dependence. It arises from the truncation of the perturbative
expansion at a fixed order in the strong coupling constant $\as$. A large
dependence on $\mu$, therefore, implies a large theoretical uncertainty. 

The scale dependence of the tree-level and Born cross section is very
similar. For the rest of this paper we will always use the Born cross
section (i.e. the leading order hard partonic cross section integrated
over the next-to-leading order parton densities) as the leading order
result. As we will see, the scale dependence is substantially reduced once
the next-to-leading order corrections are included. 

We will always consider proton--proton scattering with $\sqrt{S} =$ 500
GeV. For the strong coupling constant $\as$ we use the standard two-loop
form with $\Lambda_{\sss QCD}$ set to the value used in the parton
distribution function under consideration.  
Our default choice for the scale is 
\beq
\mu_0 \equiv \frac{1}{2}\sum_i k_{i{\sss T}},
\eeq
where the sum is over all final-state partons and $k_{i{\sss T}}$ is the
transverse momentum of parton $i$. We will study the scale
dependence of several observables by comparing the results obtained with
$\mu=\mu_0$ with those obtained with $\mu=\mu_0/2$ and $\mu=2\mu_0$, in
both the polarized and unpolarized cases. For this purpose, we fix the
parton distribution functions: our default choices are MRST~\cite{mrst} 
and GRSV~std \cite{grsv} for the unpolarized and the polarized case
respectively. We verified that our conclusions are unchanged when using 
other density sets.

\begin{figure}
\centerline{
   \epsfig{figure=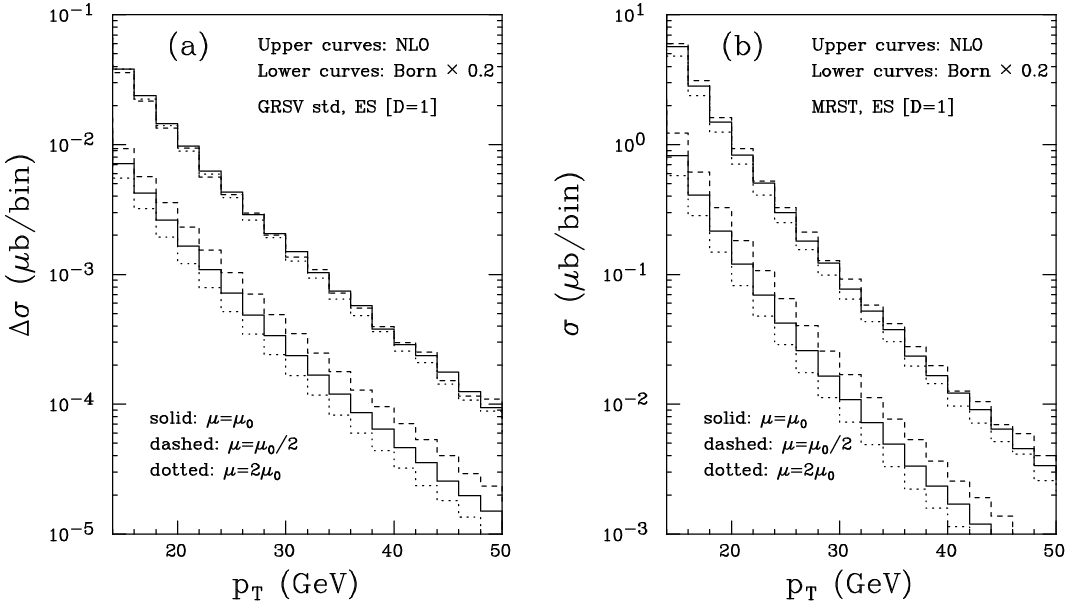,width=0.72\textwidth,clip=} }
\ccaption{}{ \label{figPT}
   Scale dependence of the next-to-leading order and Born 
   $p_{\sss T}$-distributions 
   for the Ellis--Soper algorithm with $D=1$. (a) Polarized $pp$ scattering
   and (b) unpolarized $pp$ scattering at $\sqrt{S} =$ 500 GeV. The range
   of the pseudo-rapidity is  restricted to $|\eta| < 1 $.
}
\end{figure}                                                              

To start with, we consider the inclusive $p_{\sss T}$-distributions, where
$p_{\sss T}$ is the transverse momentum of the jet. As in
fig.~\ref{figLO_BORN}, we use the cluster jet algorithm  with the
resolution parameter $D=1$ and require $|\eta| < 1 $. We checked that,
basically, the definition of a jet through a cone algorithm only amounts
to a change in the normalization. Concerning the scale dependence, the
different jet definitions give very similar results.  In fig.~\ref{figPT}a
we show  
the next-to-leading and leading order distributions in the polarized case
for the three different scales: $\mu_0, 2 \mu_0 $ and $\frac{1}{2} \mu_0 $.
Note that the Born results have been rescaled in order to disentangle the
two sets of curves. Figure~\ref{figPT}b is the corresponding plot for the
unpolarized case. Clearly, the dependence on the scale is substantially
reduced when going to next-to-leading order. The situation in the
polarized case is indeed very similar to the unpolarized  one.  

We also investigated the distribution in pseudo-rapidity and verified that
the scale dependence is reduced in a similar manner as for the
$p_{\sss T}$-distribution. Again, the dependence on the scale is very
similar to the unpolarized case. 

\begin{figure}
\centerline{
   \epsfig{figure=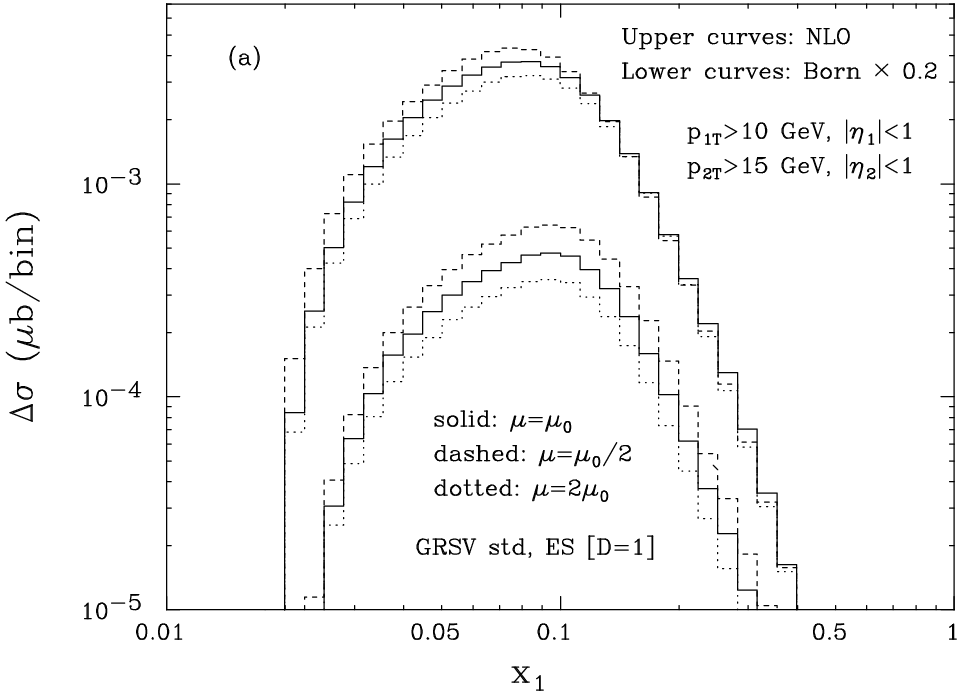,width=0.48\textwidth,clip=}
   \hfill
   \epsfig{figure=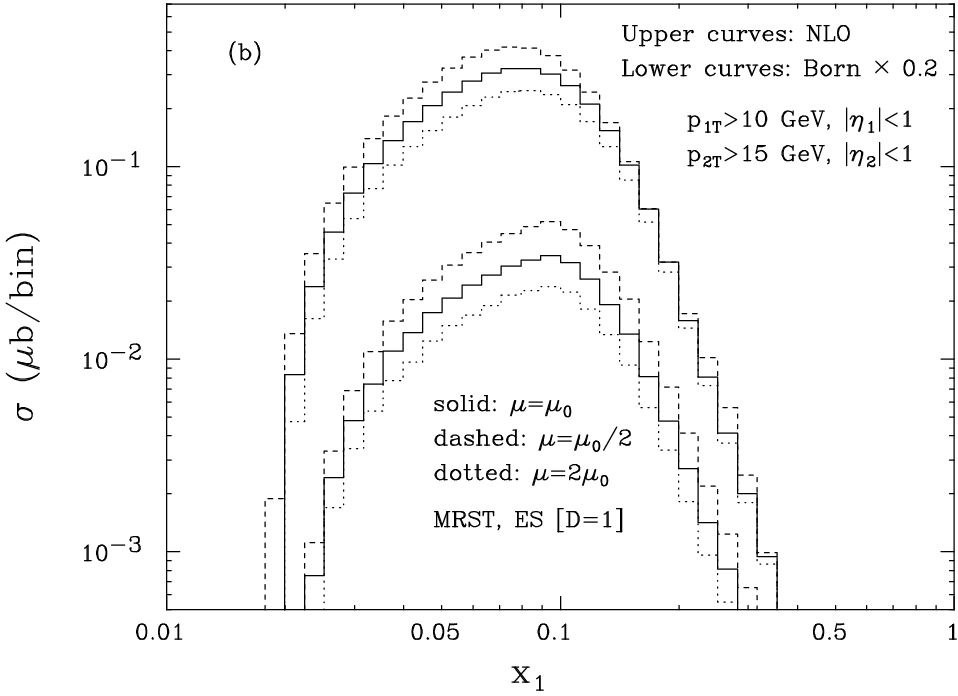,width=0.48\textwidth,clip=} }
\ccaption{}{ \label{figX1}
   As in fig.~\ref{figPT}, for the $x_1$-distribution (see the text for
   the definition).  
}
\end{figure}                                                              

A more stringent test on the perturbative stability of fixed-order
QCD calculations can be made by considering quantities that are
more exclusive than single-inclusive ones. At next-to-leading order,
this basically means double-differential cross sections. Although
the definition of these quantities is to some extent arbitrary,
we will use the most commonly adopted prescription: we select 
all the events with two or more jets, we apply suitable cuts to
the two leading jets (i.e. those with the largest transverse 
momentum) of each event, and we finally compute correlations between 
these two leading jets. To be specific, we require
\beq
p_{1{\sss T}}>p_{1{\sss T}}^{cut},\;\;\;\;\abs{\eta_1}<1,\;\;\;\;
p_{2{\sss T}}>p_{2{\sss T}}^{cut},\;\;\;\;\abs{\eta_2}<1,
\label{DDcutsdef}
\eeq
where $p_{i{\sss T}}$ and $\eta_i$ denote the transverse momenta and the
pseudo-rapidities of the two leading jets. We considered the two cases
\beq
p_{1{\sss T}}^{cut}=p_{2{\sss T}}^{cut}=10~{\rm GeV},
\label{symmcut}
\eeq
and
\beq
p_{1{\sss T}}^{cut}=10~{\rm GeV},\;\;\;\;\;\;
p_{2{\sss T}}^{cut}=15~{\rm GeV},
\label{asymmcut}
\eeq
and studied various correlations, such as the angular distance in the
plane perpendicular to the beam axis $\Delta\phi_{jj}$, the invariant 
mass $M_{jj}$ and the transverse momentum $p_{\sss T}^{jj}$ of the pair, 
the difference in pseudo-rapidity of the two jets $\Delta\eta_{jj}$
and so on. In the case when the cuts given in eq.~(\ref{asymmcut})
are applied, we find that the next-to-leading order results
are more stable (with respect to scale variations) than the
corresponding Born ones, where this comparison makes sense
(there are regions of the phase space that are not accessible
at the Born level. A typical example, with the cuts of eq.~(\ref{asymmcut}),
is the threshold in the invariant mass distribution. In these
regions, the NLO results display a scale dependence that is
larger than anywhere else. This is what we expect, since in these
regions our results are effectively leading order results). As an example,
we present in fig.~\ref{figX1} the distribution in the variable $x_1$, 
defined as follows
\beq
x_1 = \frac{p_{1{\sss T}}\, e^{\eta_1}  
    + p_{2{\sss T}}\, e^{\eta_2}}{\sqrt{S}}.
\eeq
The variable $x_1$ roughly corresponds to the momentum fraction of
one of the two partons of the scattering process, and it is exactly
so at the Born level. From the figure, we see that the NLO result has an 
enhanced scale dependence in the region of small $x_1$; this is due to
the fact that the small $x_1$ values correspond to almost back-to-back jets, 
that is to a configuration which is sensitive to soft gluon emission.
However, in the whole remaining range of $x_1$ the scale dependence of the
NLO result is smaller than that of the Born result.

The case when the cuts given in eq.~(\ref{symmcut}) are applied is
more problematic in the framework of a fixed-order QCD calculation.
This situation has been discussed in great detail in ref.~\cite{sg2};
here, we just remind the reader that, although the cuts given
in eq.~(\ref{symmcut}) define an infrared-safe cross section
(and can therefore be implemented without any problem by the
experiments), there are regions in the phase space (basically,
all those corresponding to an exactly back-to-back configuration
of the two leading jets) where any fixed-order QCD calculation breaks 
down, and an all-order resummation would be required. We explicitly
verified that, in the regions $\Delta\phi_{jj}\simeq\pi$,
$p_{\sss T}^{jj}\simeq 0$ and close to the threshold in $M_{jj}$,
the scale dependence of the NLO result is much larger than
that of the LO result, thus signalling a failure in the perturbative
expansion. However, we stress that, in all the remaining regions
of the phase space, the cuts of eq.~(\ref{symmcut}) result in
a cross section as well behaved as that obtained by applying
the cuts of eq.~(\ref{asymmcut}).

Finally we would like to mention that we have computed all the
aforementioned distributions by using both the ES and the cone jet-finding
algorithms. We observed only minor differences between the two resulting
cross sections.

\section{Dependence of Observables upon Parton Densities \label{DepPDF}}
\setcounter{equation}{0}

Up to this point, we have shown that the polarized jet cross
sections at next-to-leading order display a remarkable stability under
scale changes in the upper end of the energy range probed at RHIC. It is 
therefore sensible to use our code to investigate in some detail a 
few phenomenological issues relevant to hadronic physics at RHIC.

We thus turn to the problem of studying the dependence of the
theoretical predictions upon the perturbatively non-calculable
quantities that enter eq.~(\ref{factth1}), namely the
parton densities and the value of $\Lambda_{\sss QCD}$.
We will not perform an analysis of the separate dependence
of our results upon these two quantities, since the polarized
parton density parametrizations are only available with a single 
value of $\Lambda_{\sss QCD}$. In the following, $\Lambda_{\sss QCD}$ 
will therefore always be set equal to the value associated
to the various parton density sets used. In this section, our
predictions are obtained by defining the jets with the ES
algorithm with $D=1$, and the scales have been set to the
default value $\mu=\mu_0$. 

\begin{figure}
\centerline{
   \epsfig{figure=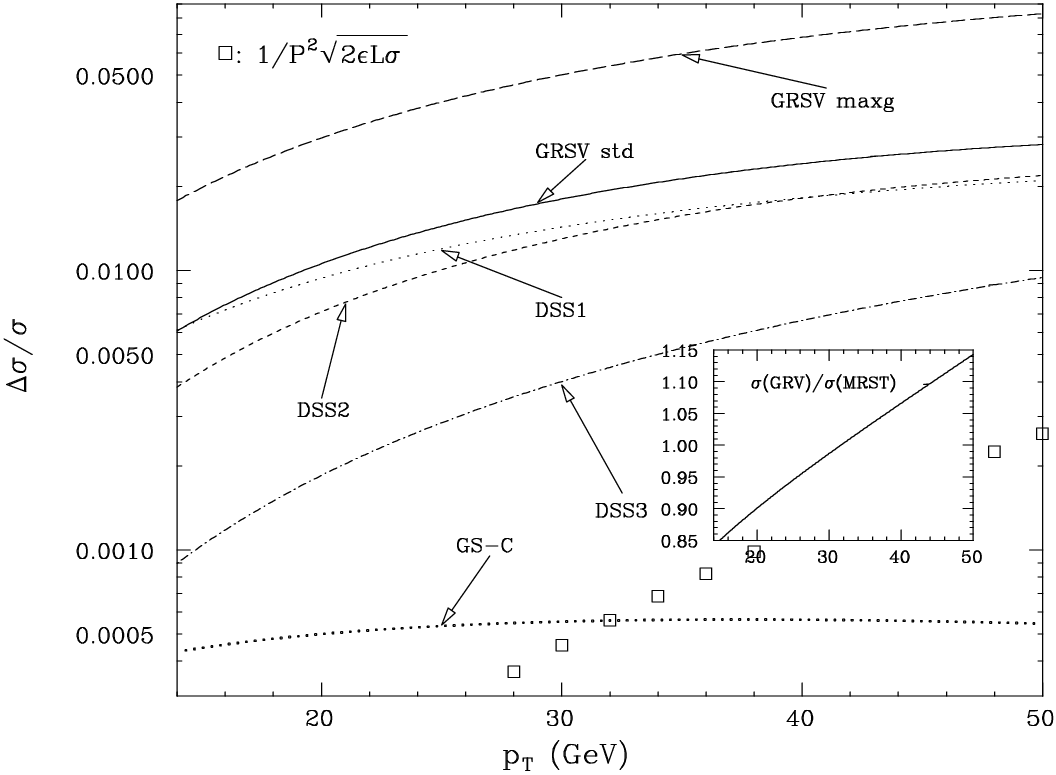,width=0.64\textwidth,clip=} }
\ccaption{}{ \label{figAS_PT}
Asymmetry versus transverse momentum for various
polarized parton densities. The minimum measurable value for the
asymmetry is also shown.
}
\end{figure}

We start by considering the asymmetry
\beq
{\cal A}_{p_{\sss T}}=\frac{d\Delta\sigma/dp_{\sss T}}{d\sigma/dp_{\sss T}}.
\label{asypt}
\eeq
Here $\Delta\sigma$ and $\sigma$ are the one-jet inclusive cross
sections for polarized and unpolarized scattering respectively,
and $p_{\sss T}$ is the transverse momentum of the observed jet. A cut
$\abs{\eta}<1$ has also been applied. In fig.~\ref{figAS_PT}, 
${\cal A}_{p_{\sss T}}$ is shown  as a function of $p_{\sss T}$. The
results for $\Delta\sigma$ have been obtained by
choosing the six different parametrizations of the polarized parton
densities previously mentioned. The unpolarized cross section $\sigma$ has
been evaluated using the MRST set. Figure~\ref{figAS_PT} clearly shows 
that the choice of the polarized parton densities induces an 
uncertainty on the theoretical results of more than two orders of 
magnitude. This enormous spread is basically due to the fact
that the polarized gluon density is very poorly constrained by
present DIS data, and at this energy the jet cross section is
dominated by $gg$- and $qg$-initiated parton processes.
Therefore, there is a chance that the measurement of the polarized
jet cross section at RHIC will be useful in order to rule out
some of the polarized sets that are at present consistent with
the data. However, since we see from fig.~\ref{figAS_PT} that the
asymmetry is always rather small, regardless of the specific
densities used, the statistics will have to be large.
To estimate the minimum value of the asymmetry observable 
at RHIC, we use the well-known formula
\beq
\left({\cal A}_{p_{\sss T}}\right)_{min}=
\frac{1}{P^2} \frac{1}{\sqrt{2\sigma {\cal L}\ep}} ,
\label{Amin}
\eeq
where ${\cal L}$ is the integrated luminosity, $P$ is the polarization of
the beam, and the factor $\ep\le 1$ accounts for experimental efficiencies;
$\sigma$ is the unpolarized cross section integrated over a small
range in transverse momentum ($p_{\sss T}$ bin). The quantity
defined in eq.~(\ref{Amin}) is plotted (boxes) in fig.~\ref{figAS_PT},
for $\ep=1$, $P=0.7$, ${\cal L}=100$~pb$^{-1}$ and a $p_{\sss T}$-bin size
of 2 GeV. From the figure we see that the asymmetry
defined in eq.~(\ref{asypt}) is measurable if the polarized parton
densities are as described by most of the sets considered here.
On the other hand, if the densities are as suggested by the GS-C
set, the measurement
will be possible only by combining a large statistics with a good
overall efficiency (clearly, it is also possible to enlarge the 
bin size, in this way decreasing the minimum observable asymmetry,
at the price of losing resolution). As a general feature, we may notice 
that the slope of the minimum observable asymmetry is steeper than the
slope of the theoretically predicted asymmetries, increasing with 
increasing transverse momentum. This means that, although the value of the
asymmetry is larger at high $p_{\sss T}$ than at small $p_{\sss T}$,
the measurement in this region will be more problematic.
We also point out that, apart from the case of GS-C, the shape
of the asymmetries obtained with the different parton density
sets is rather similar, which just reflects the fact that the various gluon
densities are rather similar in shape. 

We can still argue that the values for the asymmetry displayed
in fig.~\ref{figAS_PT} are artificially small, because of the
different values of $\Lambda_{\sss QCD}$ used when computing 
$\Delta\sigma$ and $\sigma$. In the latter case, the MRST
set has $\lambdamsb=220$~MeV, while in the former case the 
typical value is around $\lambdamsb=130$~MeV. For this reason,
in the inset of fig.~\ref{figAS_PT} we plot the ratio of the
unpolarized cross sections obtained with the MRST and the GRV sets
(the GRV set has $\lambdamsb=131$~MeV). We see that if we had plotted
the asymmetries using GRV instead of MRST, the final result would differ 
by 15\% at the most, with respect to what is shown in fig.~\ref{figAS_PT}. 
Our previous conclusions are therefore unaffected by the choice of the
unpolarized densities. Finally, we verified that, if one defines the jets
using a cone algorithm instead of the ES algorithm, the results are
practically unchanged: as one might have anticipated, the details of the
jet definition `cancel' in the ratio that defines the asymmetry.

\begin{figure}
\centerline{
   \epsfig{figure=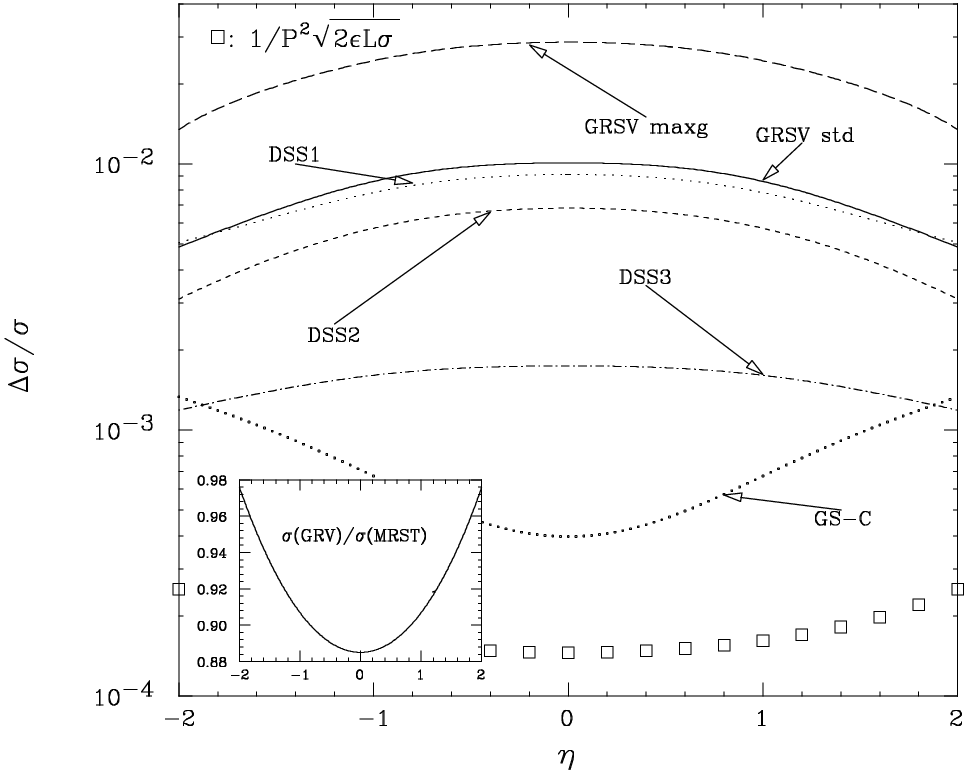,width=0.64\textwidth,clip=} }
\ccaption{}{ \label{figAS_ETA}
Asymmetry versus pseudo-rapidity for various
polarized parton densities. The minimum measurable value for the
asymmetry is also shown.
}
\end{figure}

The definition given in eq.~(\ref{asypt}) can be generalized to
any single-inclusive or double-differential observable.
In fig.~\ref{figAS_ETA} we plot the asymmetry for single-inclusive jet
production, as a function of the pseudo-rapidity 
of the jet, with a cut \mbox{$p_{\sss T}>15$~GeV}. As should
be clear from the small-$p_{\sss T}$ region of fig.~\ref{figAS_PT},
also in this case all the predictions obtained with
different density sets lie above the minimum observable asymmetry 
calculated with $\ep=1$ and ${\cal L}=100$~pb$^{-1}$ (the bin
size is 0.2). As in the case of transverse momentum, all
the shapes are rather similar, except for the one relevant to
GS-C, which has a local minimum at $\eta=0$. The study of the
asymmetry in the region around $\eta=0$ could therefore be used 
to infer information on the shape of the polarized gluon density.
We verified again that fig.~\ref{figAS_ETA} is unchanged if
we use a jet-finding algorithm based upon a cone prescription.

We also computed single-inclusive asymmetries at the Born level.
The results differ from those presented in fig.~\ref{figAS_PT}
and fig.~\ref{figAS_ETA} for a factor up to 20\%. The shape is
also different. Therefore, NLO corrections give non-trivial information
on the structure of the asymmetries.

\begin{figure}
\centerline{
   \epsfig{figure=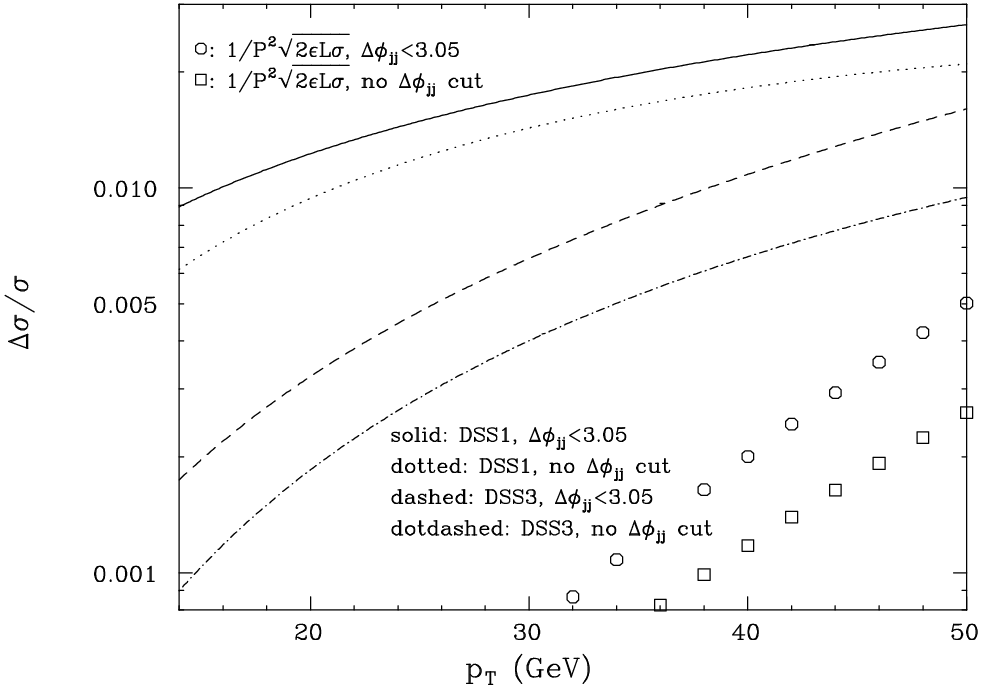,width=0.64\textwidth,clip=} }
\ccaption{}{ \label{figAS_PT_CUT}
Asymmetry versus transverse momentum. The effect
of a cut on the azimuthal distance between the two leading jets
of the event is shown.
}
\end{figure}

The possibility remains that, in some regions of the phase space,
the asymmetry is larger than that shown in figs.~\ref{figAS_PT}
and \ref{figAS_ETA}, and therefore that the measurement of 
the polarized cross section turns out to be easier. In the following we
give a simple example. By inspection of our results
for double-differential observables, we noticed that the $\Delta\phi_{jj}$
correlation is steeper in the case of unpolarized collisions than 
in the case of polarized ones. In other words, the
fraction of events with $\Delta\phi_{jj}\simeq\pi$ is higher in
unpolarized collisions with respect to polarized collisions. This 
means that, by rejecting all the events with $\Delta\phi_{jj}\simeq\pi$,
the asymmetry will be enhanced. We therefore considered all
the events with at least two jets, and we evaluated the azimuthal
distance $\Delta\phi_{jj}$ between the two leading jets. We
eventually rejected all the events with $\Delta\phi_{jj}>3.05$.
For all the jets in the events passing this cut, we then computed
the asymmetry as given in eq.~(\ref{asypt}). The results are
shown in fig.~\ref{figAS_PT_CUT}, for two different parton
density sets, superimposed on the results already shown
in fig.~\ref{figAS_PT} obtained with the same parton densities.
As we expected, the value of the asymmetry is indeed increased,
by a factor of about 1.5 to 2. However, if we compute the
minimum observable asymmetry as indicated before, we see that
this quantity also increased (since the unpolarized cross section
has decreased by effect of the $\Delta\phi_{jj}$ cut).
It follows that the measurement of the asymmetry in the case of
a $\Delta\phi_{jj}$ cut will be as difficult as in the case when
no cut is applied (and, in the high-$p_{\sss T}$ region, even
more difficult). On the other hand, such a possibility is still 
interesting in the sense that it can be used as a valuable test 
of the predictions of perturbative QCD, since the observable shown
in fig.~\ref{figAS_PT_CUT} is more exclusive than the one
presented in fig.~\ref{figAS_PT}.

\begin{figure}
\centerline{
   \epsfig{figure=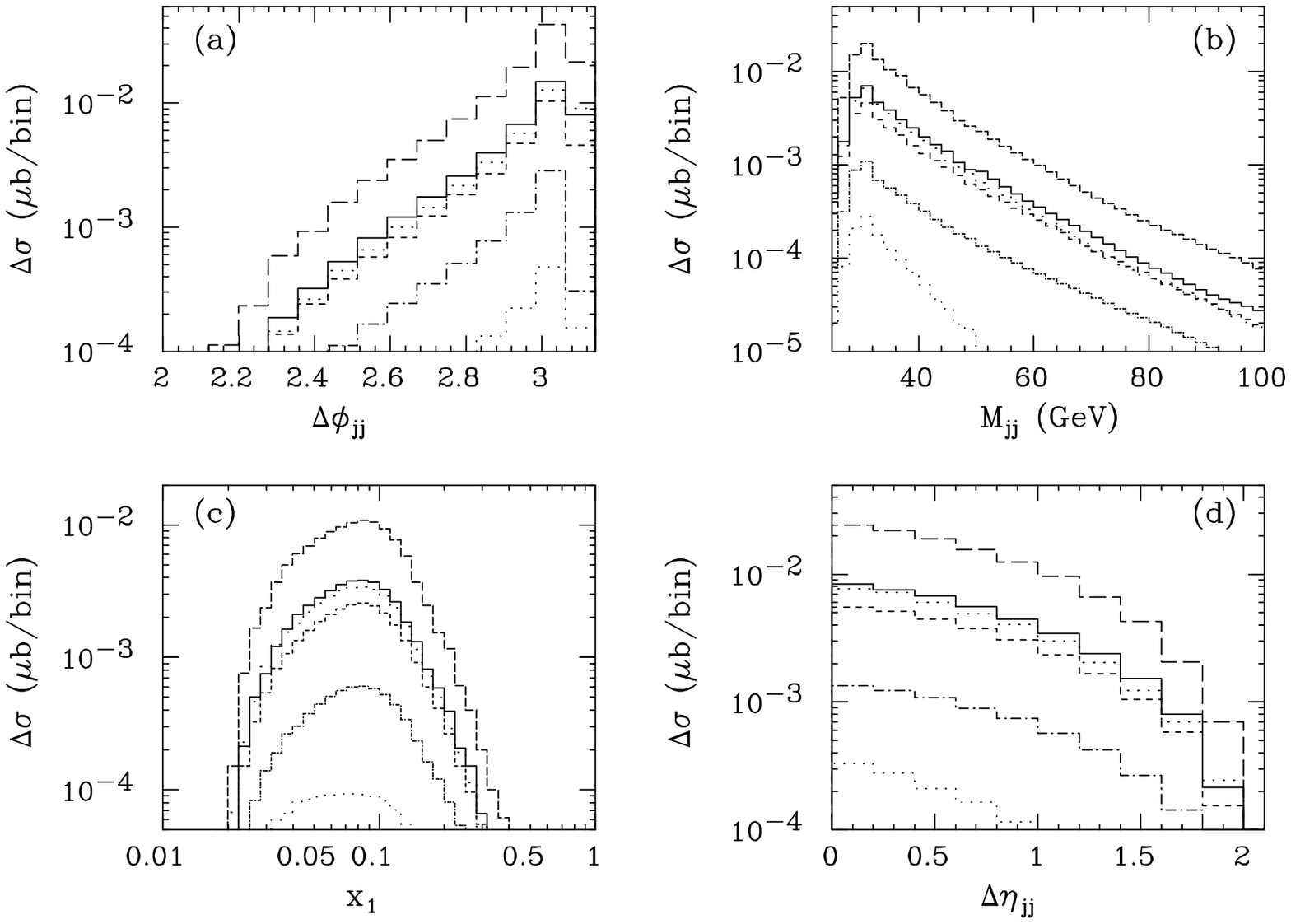,width=0.72\textwidth,clip=} }
\ccaption{}{ \label{figDOUB_DIFF}
   Double differential observables in polarized collisions
   for various parton densities. The order of the cross sections is the
   same as in fig~\ref{figAS_ETA}.
}
\end{figure}

We finally consider some double differential observables, namely
$\Delta\phi_{jj}$, $M_{jj}$, $\Delta\eta_{jj}$ and the $x_1$-distribution,
defined in the previous section; the results are presented in terms of
$\Delta\sigma$ in fig.~\ref{figDOUB_DIFF}. For the plots shown in that
figure we choose the cuts as given in eqs.~(\ref{DDcutsdef}) and
(\ref{asymmcut}). Each observable is computed with the six 
different parton density sets used previously in the case of
asymmetries. We see that, for $\Delta\phi_{jj}$, $x_1$ and
$\eta_{jj}$, the shape of the distributions is rather insensitive
to the choice of the densities. A somewhat more pronounced
dependence is displayed in the case of the invariant mass
$M_{jj}$. In fact, the typical value for the Bjorken $x$ probed
in two-jet events is of the order of \mbox{$M_{jj}/\sqrt{S}$}.
We explicitly verified that the pattern shown in fig.~\ref{figDOUB_DIFF}
is not modified if we define the jets with a cone algorithm.
The same is also true if we change the scales from their default
values, except for the $\Delta\phi_{jj}\simeq\pi$ region, where
the scale dependence is very large and the shape of the distribution
is sizeably modified. This is due to the well-known fact that
this region is particularly sensitive to the emission of quasi-soft
gluons, and a resummation to all orders would be needed to get
a reliable theoretical prediction. As discussed in section~\ref{PertStab}
we do not have a  similar problem in the small-$M_{jj}$ region, because of
the asymmetric cuts in the transverse momenta of the two leading jets. 

\section{Conclusions \label{conclusions}}
\setcounter{equation}{0}

In this paper, we have presented the first complete calculation
at next-to-leading order in perturbative QCD of one-jet and two-jet 
cross sections in polarized hadronic collisions. To this end, we
extended the general formalism of ref.~\cite{FKS}, based upon the
subtraction method and relevant for unpolarized incoming beams,
to the case of polarized beams. The resulting formulae together with the
matrix elements of refs.~\cite{MatEl,wu} can be
easily implemented in a Monte Carlo code. We used the code presented
in ref.~\cite{Jets97} for unpolarized hadron--hadron collisions,
and we suitably modified it in order to deal with the polarized
case\footnote{The codes for both the polarized and unpolarized
collisions are available upon request.}. The program outputs the 
momenta of the final-state partons plus a weight. The momenta are 
eventually used in an analysis routine to define the physical 
observables. During the same computer run, we can
obtain predictions for an arbitrary number of one-jet and two-jet
quantities, using several jet-finding algorithms, as well as 
predictions for other infrared-safe observables, such as shape variables. 
We stress that, in spite of these features, the code is not
equivalent to a Monte Carlo parton shower generator, being the
result of a fixed-order perturbative QCD calculation.

Using the aforementioned code, we investigated in some detail
the phenomenological implications of jet production at RHIC
(polarized $pp$ collisions with a maximum centre-of-mass energy
of 500~GeV). We first studied the scale dependence of single-inclusive
and double-differential jet observables, in order to determine
whether the next-to-leading order results are reliable enough to
give sensible theoretical predictions. We found that the 
next-to-leading order corrections sizeably reduce the scale 
dependence with respect to the Born results. Furthermore, the polarized
next-to-leading order cross sections display a scale dependence 
comparable to that of the corresponding unpolarized cross sections.
We then turned to the study of the dependence of our results
upon the choice of the polarized parton densities. These quantities
are very poorly constrained by the available DIS data at present,
especially for the gluon density, which enters the dominant
contributions to the jet cross sections in the RHIC energy range.
It follows that, depending upon the specific parton density set
adopted, our predictions can vary by about two orders of magnitude.
This fact in turn implies that measurements of jet observables
at RHIC will be helpful in constraining the polarized densities.
In fact, we have shown that, if the design integrated luminosity 
will be obtained, a detailed study of the polarized jet cross
sections will be possible.

\section*{ Acknowledgements}

We warmly acknowledge J.Ph. Guillet for his collaboration at an early
stage of this work.  The work of D.deF. was partially supported by the
World Laboratory (project T1). This work was supported in part  by the EU
Fourth Framework Programme `Training and Mobility of Researchers', Network
`Quantum Chromodynamics and the Deep Structure of Elementary Particles',
contract FMRX-CT98-0194 (DG 12-MIHT). 

\section*{Appendix:
  Parton Generator for Polarized Hadron--Hadron Scattering }
  \def\thesection{A}
  \def\theequation{A.\arabic{equation}}

We will briefly describe here the main differences between parton
generators for unpolarized and polarized hadron--hadron scattering. A
detailed description of the method can be found in ref.~\cite{Jets97};
here we only explicitly present the steps necessary to convert the parton
generator of ref.~\cite{Jets97} into a parton generator relevant to
polarized collisions. 

As already mentioned, a fake measurement function is used to disentangle
the various infrared singularities appearing in the phase space,
in order to apply the subtraction method to terms that contain
one soft and one collinear singularity at most. In this respect,
there is no difference between the cases of polarized and unpolarized
scattering. Furthermore, the proof of ref.~\cite{FKS}, that the 
singularities which appear in the intermediate steps of the calculation 
eventually cancel in the sum that defines physically observable quantities, 
also goes unchanged. This is due to the fact that, as in the unpolarized case, 
the singular part of a matrix element has the form of a reduced matrix 
element convoluted with a universal (i.e. process-independent) kernel. 

We start from eq.~(\ref{factth1}); as in ref.~\cite{Jets97}, we 
will deal in the following with the case of $N-1$ jet production.
We write the polarized partonic cross section at NLO as
\beq
d\hat{\sigma}_{\aoat}(\lambda_1;\lambda_2)=
d\hat{\sigma}_{\aoat}^{(0)}(\lambda_1;\lambda_2)
+d\hat{\sigma}_{\aoat}^{(1,N)}(\lambda_1;\lambda_2)
+d\hat{\sigma}_{\aoat}^{(1,N-1)}(\lambda_1;\lambda_2),
\label{NLOres}
\eeq
where $d\hat{\sigma}_{\aoat}^{(0)}$ is the Born contribution. All the 
terms in the RHS of eq.~(\ref{NLOres}) are {\it finite};
\mbox{$d\hat{\sigma}_{\aoat}^{(1,N)}$} 
(\mbox{$d\hat{\sigma}_{\aoat}^{(1,N-1)}$}) corresponds to configurations
with $N$ ($N-1$) partons in the final state.

The $N$-parton contribution reads (see eq.~(A.1) of ref.~\cite{Jets97})
\beq
d\hat{\sigma}_{\aoat}^{(1,N)}(\lambda_1;\lambda_2)=
\sum_{i=3}^{N+2}\left(
d\sigma_{\aoat,i}^{(in,f)}(\lambda_1;\lambda_2)
+\sum_{\stackrel{j=3}{j\neq i}}^{N+2} 
d\sigma_{\aoat,ij}^{(out,f)}(\lambda_1;\lambda_2)\right).
\label{Nbodyxsec}
\eeq
The quantities \mbox{$d\sigma_{\aoat,i}^{(in,f)}(\lambda_1;\lambda_2)$}
and \mbox{$d\sigma_{\aoat,ij}^{(out,f)}(\lambda_1;\lambda_2)$} can
be obtained from eqs.~(A.3) and~(A.11) of ref.~\cite{Jets97}
respectively, with the formal substitution of the matrix elements
squared with their polarized counterpart
\beq
\MN(\FLNfullj)\,\longrightarrow\,\MN(\FLNfullj;\lambda_1;\lambda_2).
\label{MNbody}
\eeq
Notice that in the unpolarized case the quantity $\MN$ also includes an
average factor for the sum over the polarizations of the incoming partons,
which must be removed in the present case.

The $(N-1)$-parton contribution reads (see eq.~(A.15) of ref.~\cite{Jets97})
\beq
d\hat{\sigma}_{\aoat}^{(1,N-1)}(\lambda_1;\lambda_2)=
d\hat{\sigma}_{\aoat}^{(1,N-1v)}(\lambda_1;\lambda_2)
+d\hat{\sigma}_{\aoat}^{(1,N-1r)}(\lambda_1;\lambda_2).
\label{Nmobodyxsec}
\eeq
The quantity \mbox{$d\hat{\sigma}_{\aoat}^{(1,N-1v)}(\lambda_1;\lambda_2)$}
can be obtained from eqs.~(A.16), (A.23), (A.24) and~(A.25) of 
ref.~\cite{Jets97}, with formal substitutions analogous to that
of eq.~(\ref{MNbody})
\beqn
\MNmo(\FLNmofullj)&\longrightarrow&
\MNmo(\FLNmofullj;\lambda_1;\lambda_2),
\label{Mborn}
\\
\MNmoij(\FLNmofullj)&\longrightarrow&
\MNmoij(\FLNmofullj;\lambda_1;\lambda_2),
\label{Mij}
\\
\MNmoV_{\sss NS}(\FLNmofullj)&\longrightarrow&
\MNmoV_{\sss NS}(\FLNmofullj;\lambda_1;\lambda_2).
\label{Mvirt}
\eeqn
We stress the fact that the coefficients ${\cal Q}(\FLNmofullj)$ and
${\cal I}_{ij}^{(reg)}$, which appear in eq.~(A.16) of ref.~\cite{Jets97},
are not modified in the polarized case. The reason is the following:
${\cal Q}$ gets contributions from the final-state collinear configurations,
which are summed over polarizations, and from the soft part ($z=1$) of
the Altarelli--Parisi kernels relevant to initial-state collinear
configurations, which is the same in the polarized and unpolarized
cases. ${\cal I}_{ij}^{(reg)}$ is obtained by integration of the
eikonal factors, which appear in the soft limit of the $N$-parton
matrix element squared in both the polarized and unpolarized cases.
The information on the polarizations of the incoming partons is
therefore fully contained in Born matrix elements squared
(eq.~(\ref{Mborn})), in the colour-linked Born matrix elements squared 
(eq.~(\ref{Mij})), and in the finite part of the virtual contribution
(eq.~(\ref{Mvirt})). The colour-linked matrix elements can be defined 
exactly as in ref.~\cite{KS}, using Born amplitudes summed only over 
the helicities of the final-state partons.

The quantity \mbox{$d\hat{\sigma}_{\aoat}^{(1,N-1r)}(\lambda_1;\lambda_2)$}
in the polarized case is only slightly more complicated than in the
unpolarized case, eq.~(A.26) of ref.~\cite{Jets97}. We get
\beqn
d\hat{\sigma}_{\aoat}^{(1,N-1r)}(K_1,\lambda_1;K_2,\lambda_2)&=&
\frac{\as}{4\pi}\sum_d\int d\xi\,\Oop_{da_1}(\xi)\,
d\Xi_1\sigma_{da_2}^{(0)}((1-\xi)K_1;K_2,\lambda_2)
\nonumber \\*
&+&\frac{\lambda_1}{\abs{\lambda_1}}\frac{\as}{4\pi}
\sum_d\int d\xi\,\Delta\Oop_{da_1}(\xi)\,
d\Delta_1\sigma_{da_2}^{(0)}((1-\xi)K_1;K_2,\lambda_2)
\nonumber \\*
&+&\frac{\as}{4\pi}\sum_d\int d\xi\,\Oop_{da_2}(\xi)\,
d\Xi_2\sigma_{a_1d}^{(0)}(K_1,\lambda_1;(1-\xi)K_2)
\nonumber \\*
&+&\frac{\lambda_2}{\abs{\lambda_2}}\frac{\as}{4\pi}
\sum_d\int d\xi\,\Delta\Oop_{da_2}(\xi)\,
d\Delta_2\sigma_{a_1d}^{(0)}(K_1,\lambda_1;(1-\xi)K_2),
\nonumber \\*
\label{sigNmoR}
\eeqn
where
\beqn
d\left(\!\!\begin{array}{c}
\Xi_1 \\ \Delta_1
\end{array}\!\!\right)
\sigma_{ab}(\lambda)&=&
d\sigma_{ab}(+;\lambda)\pm d\sigma_{ab}(-;\lambda),
\\
d\left(\!\!\begin{array}{c}
\Xi_2 \\ \Delta_2
\end{array}\!\!\right)
\sigma_{ab}(\lambda)&=&
d\sigma_{ab}(\lambda;+)\pm d\sigma_{ab}(\lambda;-).
\eeqn
The form of the operator $\Oop_{ab}$ can be read from eq.~(A.26) 
of ref.~\cite{Jets97}:
\beqn
\Oop_{ab}(\xi)&=&\xi P_{ab}^{<}(1-\xi,0)
\Bigg[\uoxic\log\frac{S\delta_{\sss I}}{2\mu^2}
+2\uoxilc\Bigg]
\nonumber \\*&-&
\xi P_{ab}^{\prime <}(1-\xi,0)\uoxic 
-K_{ab}(1-\xi),
\eeqn
where $P_{ab}^{<}(z,0)+\ep P_{ab}^{\prime <}(z,0)+{\cal O}(\ep^2)$
are the unpolarized Altarelli--Parisi kernels for $z<1$ in $4-2\ep$ 
dimensions, and $K_{ab}$ define the scheme for the unpolarized
parton densities (in the $\MSB$ scheme they are equal to zero).
We also have 
\beqn
\Delta\Oop_{ab}(\xi)&=&\xi \Delta P_{ab}^{<}(1-\xi,0)
\Bigg[\uoxic\log\frac{S\delta_{\sss I}}{2\mu^2}
+2\uoxilc\Bigg]
\nonumber \\*&-&
\xi \Delta P_{ab}^{\prime <}(1-\xi,0)\uoxic 
-\Delta K_{ab}(1-\xi),
\eeqn
where, as usual,
\beq
\Delta P_{ab}^{<}=P_{a+b+}^{<}-P_{a-b+}^{<}.
\eeq

It is now trivial to get the quantities $d\Delta\hat{\sigma}_{\aoat}$
which appear in eq.~(\ref{factth2}). From eqs.~(\ref{NLOres}),
(\ref{Nbodyxsec}) and~(\ref{Nmobodyxsec}) we get
\beqn
d\Delta\hat{\sigma}_{\aoat}&=&
d\Delta\hat{\sigma}_{\aoat}^{(0)}
+\sum_{i=3}^{N+2}\left(
d\Delta\sigma_{\aoat,i}^{(in,f)}
+\sum_{\stackrel{j=3}{j\neq i}}^{N+2} 
d\Delta\sigma_{\aoat,ij}^{(out,f)}\right)
\nonumber \\*
&+&d\Delta\hat{\sigma}_{\aoat}^{(1,N-1v)}
+d\Delta\hat{\sigma}_{\aoat}^{(1,N-1r)}.
\eeqn
The first four terms in the RHS of this equation can be obtained from
the corresponding terms in the unpolarized case with the substitution
\beq
{\cal M}\,\longrightarrow\,\Delta {\cal M},
\eeq
which directly follows from eqs.~(\ref{MNbody}),
(\ref{Mborn})--(\ref{Mvirt}). The form for
\mbox{$d\Delta\hat{\sigma}_{\aoat}^{(1,N-1r)}$} can be directly obtained
from eq.~(\ref{sigNmoR}). We have 
\beqn
d\Delta\hat{\sigma}_{\aoat}^{(1,N-1r)}(K_1;K_2)&=&
\frac{\as}{2\pi}\sum_d\int d\xi\,\Delta\Oop_{da_1}(\xi)\,
d\Delta\sigma_{da_2}^{(0)}((1-\xi)K_1;K_2)
\nonumber \\*
&+&\frac{\as}{2\pi}\sum_d\int d\xi\,\Delta\Oop_{da_2}(\xi)
d\Delta\sigma_{a_1d}^{(0)}(K_1;(1-\xi)K_2),
\eeqn
which is in fact completely analogous to eq.~(A.26) of ref.~\cite{Jets97}.

In summary, the main structure of a computer code which evaluates jet 
cross sections is identical in the unpolarized and polarized cases. 
No conceptual modification is required; in particular, there is
no need to deal explicitly with the polarizations of the incoming
hadrons/partons. The full information on the polarizations can be
embedded in the `polarized' Altarelli--Parisi kernels $\Delta P$, 
and in the `polarized' matrix elements \mbox{$\Delta {\cal M}$}. Both
quantities can be treated as black boxes, exactly like the corresponding
quantities in the unpolarized case.


\def\np#1#2#3  {{\it Nucl. Phys. }{\bf #1} (19#3) #2} 
\def\nc#1#2#3  {{\it Nuovo. Cim. }{\bf #1} (19#3) #2} 
\def\pl#1#2#3  {{\it Phys. Lett. }{\bf #1} (19#3) #2} 
\def\pr#1#2#3  {{\it Phys. Rev. }{\bf #1} (19#3) #2} 
\def\prl#1#2#3  {{\it Phys. Rev. Lett. }{\bf #1} (19#3) #2} 
\def\prep#1#2#3 {{\it Phys. Rep. }{\bf #1} (19#3) #2} 
\def\zp#1#2#3  {{\it Z. Phys. }{\bf #1} (19#3) #2} 
\def\rmp#1#2#3  {{\it Rev. Mod. Phys. }{\bf #1} (19#3) #2} 
\def\hepph#1 {hep-ph/#1}


\begin{thebibliography}{99}

\bibitem{data}
 For a compilation of references to the data, see  P.J.~Mulders and
 T.~Sloan, Summary talk of Spin Physics Working Group at the $6^{\rm th}$
 International Workshop on Deep Inelastic Scattering and  QCD, Brussels
 (1998), \hepph{9806314}. 
\bibitem{grsv} M.\ Gl\"{u}ck, E.\ Reya, M. Stratmann and W.\ Vogelsang,
  \pr{D53}{4775}{96} .  
\bibitem{gs}
T.\ Gehrmann and W.J.\ Stirling, \pr{D53}{6100}{96} .
\bibitem{dss}
 D.~de~Florian, O.A.~Sampayo and R.~Sassot, \pr{D57}{5803}{98} .
\bibitem{lead}
 C.~Bourrely {\it et al.}, \hepph{9803229};\\ 
 L.~Gordon, M.~Goshtasbpour and G.P.~Ramsey, \hepph{9803351};\\  
 E.~Leader, A.V.~Sidorov and D.B.~Stamenov,  \hepph{9807251}.
\bibitem{marco}
 M.~Stratmann, in proceedings of the `2nd Topical Workshop on Deep
 Inelastic  Scattering off Polarized Targets: Theory Meets  Experiment
 (SPIN 97)', DESY-Zeuthen, Germany, September 1997, p. 94.,
 \hepph{9710379}. 
\bibitem{bfr}
 R.D.~Ball, S.~Forte and G.~Ridolfi, \pl{B378}{255}{96} ; \\ 
 G.~Altarelli, R.D.~Ball, S.~Forte and G.~Ridolfi,  \np{B496}{337}{97} ; \\
 G.~Altarelli, R.D.~Ball, S.~Forte and G.~Ridolfi, Talks given at
 Cracow Epiphany Con\-ference on Spin Effects in Particle Physics and
 Tempus Workshop, Cracow, Poland, 9-11 Jan 1998. 
 {\it Acta Phys.Polon.} {\bf B29} (1998) 1145, \hepph{9803237}. 
\bibitem{soffer}
 J.~Soffer and J.M.~Virrey, \np{B509}{297}{98} .
\bibitem{MatEl}
 Z.~Bern and D.A.~Kosower, \np{B379}{451}{92} ; \\
 Z.~Kunszt, A.~Signer and Z.~Tr\'ocs\'anyi, \np{B411}{397}{94} . 
\bibitem{wu}
 R.\ Gastmans and T.T.\ Wu, {\it The Ubiquitous Photon}, 1990, 
 Clarendon Press, Oxford.
\bibitem{GGK}
 W.T.~Giele and E.W.N.~Glover, \pr{D46}{1980}{92} ;\\
 W.T.~Giele, E.W.N.~Glover and D.A.~Kosower, \np{B403}{633}{93} .
\bibitem{FKS}
 S.~Frixione, Z.~Kunszt and A.~Signer, \np{B467}{399}{96} .
\bibitem{CS}
 S.~Catani and M.~Seymour, \np{B485}{291}{97} 
  (Erratum \np{B510}{503}{98} ).
\bibitem{Jets97}
 S.~Frixione, \np{B507}{295}{97}  .
\bibitem{CSS}
 J.C.~Collins, D.E.~Soper and G.~Sterman, in {\it Perturbative
 Quantum Chromodynamics}, ed. A.~Mueller, 1989, World Scientific,
 Singapore, and references therein.
\bibitem{EScluster}
 S.~Ellis and D.E.~Soper, \pr{D48}{3160}{93} .
\bibitem{grv}
M.\ Gl\"uck, E.\ Reya and A.\ Vogt, 
\zp{C67}{433}{95} .
\bibitem{mrst}
 A.D.~Martin, R.G.~Roberts, W.J.~Stirling and R.S.~Thorne, {\it
   Eur. Phys. J.} {\bf C4} (1998) 463. 
\bibitem{sg2}
 S.~Frixione and G.~Ridolfi, \np{B507}{315}{97} .
\bibitem{KS}
 Z.~Kunszt and D.E.~Soper, \pr{D46}{192}{92} .
\end{thebibliography}
\end{document}